\documentstyle[twoside]{article}

\catcode`\@=11
\long\def\@makefntext#1{
\protect\noindent \hbox to 3.2pt {\hskip-.9pt
$^{{\eightrm\@thefnmark}}$\hfil}#1\hfill}		

\def\@makefnmark{\hbox to 0pt{$^{\@thefnmark}$\hss}}	

\def\ps@myheadings{\let\@mkboth\@gobbletwo
\def\@oddhead{\hbox{}
\rightmark\hfil\eightrm\thepage}
\def\@oddfoot{}\def\@evenhead{\eightrm\thepage\hfil
\leftmark\hbox{}}\def\@evenfoot{}
\def\sectionmark##1{}\def\subsectionmark##1{}}

\oddsidemargin=\evensidemargin
\newcounter{sectionc}\newcounter{subsectionc}\newcounter{subsubsectionc}
\renewcommand{\section}[1] {\vspace{12pt}\addtocounter{sectionc}{1}
\setcounter{subsectionc}{0}\setcounter{subsubsectionc}{0}\noindent
	{\tenbf\thesectionc. #1}\par\vspace{5pt}}
\renewcommand{\subsection}[1] {\vspace{12pt}\addtocounter{subsectionc}{1}
	\setcounter{subsubsectionc}{0}\noindent
	{\bf\thesectionc.\thesubsectionc. {\kern1pt \bfit #1}}\par\vspace{5pt}}
\renewcommand{\subsubsection}[1] {\vspace{12pt}\addtocounter{subsubsectionc}{1}
	\noindent{\tenrm\thesectionc.\thesubsectionc.\thesubsubsectionc.
	{\kern1pt \tenit #1}}\par\vspace{5pt}}
\newcommand{\nonumsection}[1] {\vspace{12pt}\noindent{\tenbf #1}
	\par\vspace{5pt}}

\newcounter{appendixc}
\newcounter{subappendixc}[appendixc]
\newcounter{subsubappendixc}[subappendixc]
\renewcommand{\thesubappendixc}{\Alph{appendixc}.\arabic{subappendixc}}
\renewcommand{\thesubsubappendixc}
	{\Alph{appendixc}.\arabic{subappendixc}.\arabic{subsubappendixc}}

\renewcommand{\appendix}[1] {\vspace{12pt}
        \refstepcounter{appendixc}
        \setcounter{figure}{0}
        \setcounter{table}{0}
        \setcounter{lemma}{0}
        \setcounter{theorem}{0}
        \setcounter{corollary}{0}
        \setcounter{definition}{0}
        \setcounter{equation}{0}
        \renewcommand{\thefigure}{\Alph{appendixc}.\arabic{figure}}
        \renewcommand{\thetable}{\Alph{appendixc}.\arabic{table}}
        \renewcommand{\theappendixc}{\Alph{appendixc}}
        \renewcommand{\thelemma}{\Alph{appendixc}.\arabic{lemma}}
        \renewcommand{\thetheorem}{\Alph{appendixc}.\arabic{theorem}}
        \renewcommand{\thedefinition}{\Alph{appendixc}.\arabic{definition}}
        \renewcommand{\thecorollary}{\Alph{appendixc}.\arabic{corollary}}
        \renewcommand{\theequation}{\Alph{appendixc}.\arabic{equation}}
        \noindent{\tenbf Appendix \theappendixc #1}\par\vspace{5pt}}
\newcommand{\subappendix}[1] {\vspace{12pt}
        \refstepcounter{subappendixc}
        \noindent{\bf Appendix \thesubappendixc. {\kern1pt \bfit #1}}
	\par\vspace{5pt}}
\newcommand{\subsubappendix}[1] {\vspace{12pt}
        \refstepcounter{subsubappendixc}
        \noindent{\rm Appendix \thesubsubappendixc. {\kern1pt \tenit #1}}
	\par\vspace{5pt}}

\newcommand{\textlineskip}{\baselineskip=13pt}
\newcommand{\smalllineskip}{\baselineskip=10pt}

\def\eightcirc{
\begin{picture}(0,0)
\put(4.4,1.8){\circle{6.5}}
\end{picture}}
\def\eightcopyright{\eightcirc\kern2.7pt\hbox{\eightrm c}}

\newcommand{\copyrightheading}[1]
	{\vspace*{-2.5cm}\smalllineskip{\flushleft
	{\footnotesize International Journal of Modern Physics A, #1}\\
	{\footnotesize $\eightcopyright$\, World Scientific Publishing
	 Company}\\
	 }}

\newcommand{\publisher}[2]{{\begin{center}\footnotesize\smalllineskip
	Received #1\\
	\end{center}
	}}

\def\abstracts#1#2#3{{
	\centering{\begin{minipage}{4.5in}\baselineskip=10pt\footnotesize
	\parindent=0pt #1\par
	\parindent=15pt #2\par
	\parindent=15pt #3
	\end{minipage}}\par}}

\renewenvironment{thebibliography}[1]
	{\frenchspacing
	 \ninerm\baselineskip=11pt
	 \begin{list}{\arabic{enumi}.}
	{\usecounter{enumi}\setlength{\parsep}{0pt}
	 \setlength{\leftmargin 12.7pt}{\rightmargin 0pt} 
	 \setlength{\itemsep}{0pt} \settowidth
	{\labelwidth}{#1.}\sloppy}}{\end{list}}

\newcounter{itemlistc}
\newcounter{romanlistc}
\newcounter{alphlistc}
\newcounter{arabiclistc}

\newcommand{\fcaption}[1]{
        \refstepcounter{figure}
        \setbox\@tempboxa = \hbox{\footnotesize Fig.~\thefigure. #1}
        \ifdim \wd\@tempboxa > 5in
           {\begin{center}
        \parbox{5in}{\footnotesize\smalllineskip Fig.~\thefigure. #1}
            \end{center}}
        \else
             {\begin{center}
             {\footnotesize Fig.~\thefigure. #1}
              \end{center}}
        \fi}

\newcommand{\tcaption}[1]{
        \refstepcounter{table}
        \setbox\@tempboxa = \hbox{\footnotesize Table~\thetable. #1}
        \ifdim \wd\@tempboxa > 5in
           {\begin{center}
        \parbox{5in}{\footnotesize\smalllineskip Table~\thetable. #1}
            \end{center}}
        \else
             {\begin{center}
             {\footnotesize Table~\thetable. #1}
              \end{center}}
        \fi}

\def\@citex[#1]#2{\if@filesw\immediate\write\@auxout
	{\string\citation{#2}}\fi
\def\@citea{}\@cite{\@for\@citeb:=#2\do
	{\@citea\def\@citea{,}\@ifundefined
	{b@\@citeb}{{\bf ?}\@warning
	{Citation `\@citeb' on page \thepage \space undefined}}
	{\csname b@\@citeb\endcsname}}}{#1}}

\newif\if@cghi
\def\cite{\@cghitrue\@ifnextchar [{\@tempswatrue
	\@citex}{\@tempswafalse\@citex[]}}
\def\citelow{\@cghifalse\@ifnextchar [{\@tempswatrue
	\@citex}{\@tempswafalse\@citex[]}}
\def\@cite#1#2{{$\null^{#1}$\if@tempswa\typeout
	{IJCGA warning: optional citation argument
	ignored: `#2'} \fi}}

\def\pmb#1{\setbox0=\hbox{#1}
	\kern-.025em\copy0\kern-\wd0
	\kern.05em\copy0\kern-\wd0
	\kern-.025em\raise.0433em\box0}

\def\fnt#1#2{\footnotetext{\kern-.3em
	{$^{\mbox{\scriptsize #1}}$}{#2}}}

\def\fpage#1{\begingroup
\voffset=.3in
\thispagestyle{empty}\begin{table}[b]\centerline{\footnotesize #1}
	\end{table}\endgroup}

\def\runninghead#1#2{\pagestyle{myheadings}
\markboth{{\protect\footnotesize\it{\quad #1}}\hfill}
{\hfill{\protect\footnotesize\it{#2\quad}}}}
\font\tenrm=cmr10
\font\tenit=cmti10
\font\tenbf=cmbx10
\font\bfit=cmbxti10 at 10pt
\font\ninerm=cmr9

\font\eightrm=cmr8

\def\qed{\hbox{${\vcenter{\vbox{			
   \hrule height 0.4pt\hbox{\vrule width 0.4pt height 6pt
   \kern5pt\vrule width 0.4pt}\hrule height 0.4pt}}}$}}

\begin{document}

\runninghead{A. A. Suzko  }
{Generalized Algebraic Bargmann -- Darboux Transformations $\ldots$}

\normalsize\textlineskip
\thispagestyle{empty}
\setcounter{page}{277}

\copyrightheading{			Vol. 12, No. 1 (1997) 277--282}

\vspace*{0.88truein}

\fpage{277}
\centerline{\bf GENERALIZED ALGEBRAIC}
\vspace*{0.035truein}
\centerline{\bf BARGMANN--DARBOUX TRANSFORMATIONS}
\vspace*{0.37truein}
\centerline{\footnotesize A.A.~SUZKO\footnote{
          Radiation Physics
and Chemistry Problems Institute, Academy of Sciences of Belarus, Minsk.
}}
\vspace*{0.015truein}
\centerline{\footnotesize\it Joint Institute for Nuclear Research,
Dubna, Russia }
\vspace*{0.225truein}
\publisher{14 October 1996} {}

\vspace*{0.21truein}
\abstracts{
Algebraic Bargmann and Darboux transformations for equations of a more
general form than the Schr\"odinger ones with an additional functional
dependence $h(r)$ in the right-hand side of equations are constructed.
The suggested generalized transformations turn into the Bargmann and
Darboux transformations for both fixed and variable values of energy
and an angular momentum.
}
{Key words:
Darboux transformations, Bargmann potentials, inverse scattering \\
problem.}{}

\vspace*{1pt}\textlineskip
\section{Introduction}
\vspace*{-0.5pt}
Algebraic Darboux and Bargmann transformations  essentially broaden a
class of exactly solvable problems in  quantum mechanics and, consequently,
allows one to derive solutions in a closed analytical form.
In conventional statements the potentials and pertinent solutions are
restored either at a fixed angular momentum $l$ and variable energy $E$
values or at a fixed $E$ and different $l$. In the papers \cite{1S,2S,3S}
we suggested the generalized Darboux and Bargmann transformations to
construct a wide class of potentials and appropriate solutions to the
Schr\"odinger equation for variable $l$ and energy $E$ along arbitrary
straight lines in the $(\lambda^2,E)$--plane $(\lambda=l+1/2)$. In
particular cases, when $l=const$ or $E=const$, the obtained relations
are transformed into familiar expressions for potentials and solutions
of the Bargmann type. This method was developed in \cite{3} for
constructing exactly solvable three-body models with two-central
spheroidal potentials.

The studies performed for the Schr\"odinger equations with some varying
parameters allowed us to construct algebraic Bargmann and Darboux
transformations for equations of a more general form
\begin{eqnarray}
\label{1.1}
-d^2\phi(\gamma,r)/dr^2 + V(r)\phi(\gamma,r)=\gamma^{2}h(r)\phi(\gamma,r).
\end{eqnarray}
The quantity $\gamma^2$ represents energy with a coefficient $h(r)$
dependent on the coordinate variable; the $h(r)$ should satisfy
the general requirements imposed on the potential function by the
scattering theory. These equations are applied in atomic physics,
the theory of propagation of electromagnetic waves, acoustics, geophysics,
and so on. The generalized Bargmann transformations are related to
the generalized Darboux ones and can be obtained as their superposition.
This technique of algebraic transformations is favored
because it does not use integral equations of the inverse problem and,
consequently, does not explicitly use the completeness of the set of
eigenfunctions required for its derivation; at the same time this
approach remains  a closed algebraic procedure.

\section{Darboux Transformations}
\noindent
Let us introduce the generalized Darboux transformations suggested in
\cite{echaya,lect}. The solution of Eq.(\ref{1.1}) with an unknown 
potential $V(r)$ has been sought in terms of the known solutions to 
this equation with the known potential $V^{\circ}(r)$
in the same form as for the standard Schr\"odinger equation
\begin{eqnarray}
\label{3.5}
\phi(\gamma,r)= y(r)W\{y^{\circ}(r),\phi^{\circ}(\gamma,r)\}.
\end{eqnarray}
Here $ W\{y^{\circ}(r),\phi^{\circ}(\gamma,r)\}=
y^{\circ}(r)(d\phi^{\circ}(r)/dr)-(dy^{\circ}(r)/dr)\phi^{\circ}(r) $
is the Wronskian of the functions $y^{\circ}$ and $\phi^{\circ}$.
The functions $y(r)$ and $y^{\circ}(r)$ obey Eq.(\ref{1.1})
with $V(r)$ and $V^{\circ}(r)$, respectively, at a certain fixed value
of $\gamma^2 = \gamma'^2$ that may correspond to a bound state.
Multiplying Eq.(\ref{1.1}) for $y^{\circ}(r)$
by function $\phi^{\circ}(\gamma,r)$ at arbitrary $\gamma$; whereas
Eq.(\ref{1.1}) for $\phi^{\circ}(\gamma,r)$, by $y^{\circ}(r)$, and
then subtracting the resulting expressions from one another we obtain
\begin{eqnarray}
\label{3.52}
d W(r)/dr=h(r)(\gamma'^2-\gamma^2)y^{\circ}(r)\phi^{\circ}(\gamma,r).
\end{eqnarray}
Further, let us determine the second-order derivative
$d^2\phi/dr^2$, by using Eqs.(\ref{3.5}) and (\ref{3.52}).
Then taking Eq.(\ref{1.1}) for $y(r)$ and making appropriate
transformations we get \begin{eqnarray} \label{3.52a}
{d^2\phi(\gamma,r)/ dr^2}=[V(r)-\gamma'^{2}h(r])
y(r)W\{y^{\circ}(r),\phi^{\circ}(\gamma,r)\}~~~~~~~~~~~~~~~~~~~~~~ \nonumber\\
+ 2{d [y(r)y^{\circ}(r)]\over dr} h(r)\phi^{\circ}(\gamma,r)
+ {dh(r)\over dr} y(r)y^{\circ}(r)\phi^{\circ}(\gamma,r)
+ h(r)y(r)W\{y^{\circ}(r),\phi^{\circ}(\gamma,r)\}. \nonumber
\end{eqnarray}
Using the definition (\ref{3.5}) we rewrite the latter expression in 
the form
\begin{eqnarray}
\Bigl(\frac{d^2}{dr^2}-V(r)+h(r)\gamma^2\Bigr)\phi(\gamma,r)=
2h(r)\frac{dy(r)y^{\circ}(r)}{dr}\phi^{\circ}(\gamma,r)+
y(r)y^{\circ}(r)\frac{dh(r)}{dr}\phi^{\circ}(\gamma,r).\nonumber
\end{eqnarray}
 It is clear that the condition for the right-hand side of the identity 
being zero $d(\ln y(r)y^{\circ}(r))/dr=d(\ln h(r))/dr$
makes the function $\phi(\gamma,r)$ defined by the expression
(\ref{3.5}) obey Eq.(\ref{1.1}). This condition is equivalent to
\begin{eqnarray} \label{2.4}
 y(r)= {1\over (\sqrt{h(r)}y^{\circ}(r))}.
\end{eqnarray}
Then, the solution to Eq.(\ref{1.1}) with the definition (\ref{3.5})
and at arbitrary $\gamma$ is written as
\begin{eqnarray} \label{2.5}
\phi(\gamma,r) = \frac{1}{\sqrt{h(r)}y^{\circ}(r)}
 W\{y^{\circ}(r),\phi^{\circ}(\gamma,r)\}.
\end{eqnarray}
 Now we will explicitly determine the potential $V(r)$ in terms of
the known functions $h(r),y^{\circ}(r)$ and $V^{\circ}(r)$. Using 
the relation (\ref{2.4}) in Eq.(\ref{1.1}) for the function $y(r)$
\newpage
\begin{eqnarray}
\label{3.56a}
 V(r) = \frac{d^2y(r)/dr^2}{y(r)} + h(r)\gamma'^2
= 2\Bigl(\frac{dy^{\circ}(r)/dr}{y^{\circ}(r)} \Bigr)^2
-\frac{d^{2} y^{\circ}(r)/dr^2}{y^{\circ}(r)}+ \nonumber \\
+ \frac{dy^{\circ}(r)/dr}{y^{\circ}(r)} \frac{dh(r)/dr}{h(r)}
+ h(r)\gamma'^2- \frac{1}{2} \frac{d^2h(r)/dr^2}{h(r)}
+\frac{3}{4} \Bigl(\frac{dh(r)/dr}{h(r)} \Bigr)^2 \nonumber
\end{eqnarray}
and transforming this expression on the basis of the equality
${d^{2} y^{\circ}(r)/dr^2}/{y^{\circ}(r)}=V^{\circ}(r)-h(r)\gamma'^2,$
we finally obtain the following expression for the potential:
\begin{eqnarray} \label{2.7}
V(r)= V^{\circ}(r)-2\sqrt{h(r)} \frac{d}{dr}\Bigl[\frac{1}{\sqrt{h(r)}}
\frac{d}{dr}\ln y^{\circ}(r)\Bigr]
+\sqrt{h(r)} \frac{d^2}{dr^2}\frac{1}{\sqrt{h(r)}}.
\end{eqnarray}

\section{Bargmann Transformations}
\noindent
We will look for the solution to Eq.(\ref{1.1}) in a more
general form, as compared to (\ref{3.5}),
\begin{eqnarray}
\label{3.1}
\phi(\gamma,r) = \phi^{\circ}(\gamma,r) - \sum_{\mu}^{M} y_{\mu}(r)
W\{\phi^{\circ}(\gamma_{\mu},r),\phi^{\circ}(\gamma,r)\}/
(\gamma_{\mu}^{2}-\gamma^2),
\end{eqnarray}
where
$ y_{\mu}(r)\equiv y(r,\gamma_{\mu})= C_{\mu} \phi(\gamma_{\mu},r).$
Let us now determine the conditions under which the function
$\phi(\gamma,r)$ given by (\ref{3.1})
obeys Eq.(\ref{1.1}) following a procedure analogous to that
expounded in \cite{2S}. To this end we differentiate (\ref{3.1})
twice with respect to $r$ and owing to (\ref{3.52}) we then obtain
\begin{eqnarray} \label{3.1a}
{d^2\phi(\gamma,r) \over dr^2} &=& {d^2\phi^{\circ}(\gamma,r) \over dr^2}-
\sum_{\mu}^{M}\Bigl\{{d^2 y_{\mu}(r)\over dr^2}
W\{\phi^{\circ}(\gamma_{\mu},r),\phi^{\circ}(\gamma,r)\}
/(\gamma_{\mu}^{2}-\gamma^2) \nonumber \\
&+& y_{\mu}(r){d [h(r)\phi^{\circ}(\gamma_{\mu},r)\phi^{\circ}(\gamma,r)] \over dr}
+ 2{d y_{\mu}(r)\over dr} h(r)\phi^{\circ}(\gamma_{\mu},r)\phi^{\circ}(\gamma,r)\Bigr\} \nonumber
\end{eqnarray}
and transform it with $y_{\mu}(r)$ given by (\ref{1.1}) and the
definition (\ref{3.1})
\begin{eqnarray} \label{3.1b}
\Bigl(\frac{d^2}{dr^2} - V(r) + h(r)\gamma^2 \Bigr)\phi(\gamma,r) &=&
[-V(r) + V ^{\circ}(r)]\phi^{\circ}(\gamma,r) \nonumber \\
- 2\sum_{\mu}^{M}\Bigl\{h(r)\frac{dy_{\mu}(r)\phi^{\circ}(\gamma_{\mu},r)}{dr}
&+& y_{\mu}(r)\phi^{\circ}(\gamma_{\mu},r)\frac{dh(r)}{dr}\Bigr\} \phi^{\circ}(\gamma,r).
\end{eqnarray}
The function $\phi(\gamma,r)$ satisfies Eq.(\ref{1.1}) provided the 
right-hand side of the latter relation vanishes, which is equivalent to
\begin{eqnarray} \label{3.64}
V(r) =  V^{\circ}(r)
-\sum_{\mu}^{M}\Bigl\{2 h(r)\frac{dy_{\mu}(r)\phi^{\circ}(\gamma_{\mu},r)}{dr}
+y_{\mu}(r)\phi^{\circ}(\gamma_{\mu},r)\frac{dh(r)}{dr}\Bigr\}.
\end{eqnarray}
Using the connection $y_{\mu}(r)= C_{\mu} \phi(\gamma_{\mu},r)$, we
determine the solution $ y_{\mu}(r)$ with the
potential (\ref{3.64}) from (\ref{3.1}) as follows:
\begin{eqnarray} \label{3.65}
y_{\mu}(r) =
\sum_{\nu}^{M}C_{\nu}\phi^{\circ}(\gamma_{\nu},r)P_{\nu\mu}^{-1}(r),
\end{eqnarray}
where $P_{\mu\nu}(r)=\delta_{\mu\nu}+C_{\mu}
W\{\phi^{\circ}(\gamma_{\mu},r),\phi^{\circ}(\gamma_{\nu},r)\}
/(\gamma_{\mu}^{2}-\gamma_{\nu}^2).$
\newpage
Substituting (\ref{3.65}) into (\ref{3.64}) and (\ref{3.1}) we can express 
the potential and corresponding solutions in terms of the known function 
$h(r)$ and solutions $\phi^{\circ}(\gamma,r)$
\begin{eqnarray}
\label{3.7}
V(r)= V^{\circ}(r)-2\sqrt{h(r)} \frac{d}{dr}\Bigl[\frac{1}{\sqrt{h(r)}}
\frac{d}{dr}\ln \det P(r)\Bigr];
\end{eqnarray}
\begin{eqnarray} \label{3.8}
\phi(\gamma,r) = \phi^{\circ}(\gamma,r) - \sum_{\mu}^{M} \sum_{\nu}^{M}
C_{\nu}\phi^{\circ}(\gamma_{\nu},r)P_{\nu\mu}^{-1}(r)
{W\{\phi^{\circ}(\gamma_{\mu},r),\phi^{\circ}(\gamma,r)\}\over
(\gamma_{\mu}^{2}-\gamma^2)}.
\end{eqnarray}
After the integration of Eq.(\ref{3.52}) the Wronskian
is expressed as follows
\begin{eqnarray} \label{3.69}
W\{\phi^{\circ}(\gamma_{\mu},r),\phi^{\circ}(\gamma,r)\} =
(\gamma_{\mu}^{2} - \gamma^2 )
\int_{0(r)}^{r(b)} h(r')\phi^{\circ}(\gamma_{\mu},r')\phi^{\circ}(\gamma,r')dr'
\end{eqnarray}
With allowance of this expression, the transformation (\ref{3.1}) 
and (\ref{3.8}) can be represented in the integral form.
In particular for regular solutions, the integration limits from 
$"0"$ to $"r"$ are the same as those in the Gelfand--Levitan approach;
whereas the Jost solutions, the integration limits are as those
in the Marchenko approach, i.e., from $"r"$ to $"\infty"$.
It is now clear that $\phi$ in formulae (\ref{3.1}),
(\ref{3.65})--(\ref{3.8}) stands for any solutions,
which are, generally, arbitrary, until boundary conditions are fixed.

Relations (\ref{3.1}) or (\ref{3.8}) may be represented as superposition 
of solutions of the type (\ref{2.5}). Let us take a solution 
$\phi_1(\gamma,r)$ of Eq.(\ref{1.1}) in the form (\ref{2.5}) with $V(r)$ 
determined as (\ref{2.7}). As two linearly independent solutions of 
(\ref{1.1}) we take (see Eqs.(\ref{2.4}),(\ref{2.5})
(\ref{3.69}))
\begin{eqnarray} \label{2.4a}
 y_{1}(r)= \frac{1}{\sqrt{h(r)}y^{\circ}(r)},~~~
X_{1}(r) = \frac{1}{\sqrt{h(r)}y^{\circ}(r)}
\int_{0(r)}^{r(b)} h(r')|y^{\circ}(r')|^2dr'.
\end{eqnarray}
Now we construct a solution $\eta_{1}(r)$ of Eq.(\ref{1.1}) with $V_1(r)$
(\ref{2.7}) as a linear combination of these two solutions
\begin{eqnarray} \label{2.4b}
\eta_{1}(r) = \frac{1}{\sqrt{h(r)}y^{\circ}(r)}\Biggl[1+C
\int_{0(r)}^{r(b)} h(r')|y^{\circ}(r')|^2dr'\Biggr]=
\frac{1}{\sqrt{h(r)}y^{\circ}(r)} P(r)~.
\end{eqnarray}
Then, the solution to Eq.(\ref{1.1}) with the definition (\ref{2.5})
and at arbitrary  $\gamma^2$ can be rewritten in terms of the solutions
$\eta_{1}(r)$ and $\phi_{1}(r)$ as
\begin{eqnarray} \label{3.8a}
\phi(\gamma,r) &=&
\frac{1}{\sqrt{h(r)}\eta_{1}(r)}W\{\eta_{1}(r),\phi_{1}(\gamma,r)\}
\nonumber\\
&=& \phi^{\circ}(\gamma,r) - {Cy^{\circ}(r)
W\{y^{\circ}(r),\phi^{\circ}(\gamma,r)\}\over
P(r)(\gamma^{\prime 2}-\gamma^2)}.
\end{eqnarray}
This is the solution of Eq.(\ref{1.1}), coinciding with (\ref{3.8})
at $\mu=1$, with the potential (\ref{3.7}) in the form
\begin{eqnarray}
\label{3.7a}
V(r)=V^{\circ}(r)
-2\sqrt{h(r)}\frac{d}{dr}\Bigl[\frac{1}{\sqrt{h(r)}}
\frac{d}{dr}\ln y^{\circ}(r)\Bigr]
-2\sqrt{h(r)}\frac{d}{dr}\Bigl[\frac{1}{\sqrt{h(r)}}
\frac{d}{dr}\ln\eta_{1}(r)\Bigr] \nonumber\\
+2\sqrt{h(r)}\frac{d^2}{dr^2}\frac{1}{\sqrt{h(r)}}=
V^{\circ}(r)-2\sqrt{h(r)}\frac{d}{dr}\Bigl[\frac{1}{\sqrt{h(r)}}
\frac{d}{dr}\ln P(r)\Bigr].~~~~~~~
\end{eqnarray}

Then, it is easy to show \cite{lect} that the relations (\ref{2.5}) and 
(\ref{2.7}) as well as (\ref{3.7}) and (\ref{3.8}) are reduced to the
corresponding relations for the Darboux\cite{1S} and Bargmann \cite{2S} 
transformations for both fixed or variable values of energy and orbital 
momentum. For particular cases of a different choice of $h(r)$, we may 
consider versions of these transformations for the Coulomb forces and 
a Coulomb coupling constant.

\section{Multichannel exactly solvable models.}
\noindent
Matrix generalization of the Darboux transformation has been given
in Refs.\cite{Humi,3S} for multichannel systems of Schr\"odinger
equations. In the paper \cite{3S} matrices of  solutions and potentials
in a closed analytical form  have been constructed for variable values 
of energy and angular momentum. Above approach is generalized for 
the multichannel system of coupled equations
\begin{eqnarray}
\label{4.1}
-{d^2\phi_{\alpha \beta}(r)\over dr^2} +
\sum_{\beta'}^{N} V_{\alpha \beta'}(r)\phi_{\beta'\beta}(r)
 = \gamma^{2}_{\alpha}h(r)\phi_{\alpha \beta}(r).
\end{eqnarray}
Here $\gamma=diag(\gamma_{\alpha})$. Let the vectors of solutions 
$ \psi_{\alpha}(r) =
\sum_{\beta}^{N}\phi_{\alpha \beta}(r)c_{\beta};~
\psi^{\circ}_{\alpha}(r)=\sum_{\beta}^{N}
\phi^{\circ}_{\alpha \beta}(r)c_{\beta} $
satisfy the system
\begin{eqnarray}
\label{4.3}
-{d^2\psi_{\alpha}(r)\over dr^2} + \sum_{\beta}^{N} V_{\alpha \beta}(r)
\psi_{\alpha}(r) = \gamma'^{2}_{\alpha}h\psi_{\alpha}(r)
\end{eqnarray}
with potential matrices $V(r)$ and $V^{\circ}(r)$, respectively,
within the interval $a<r<b$, where $V(r)$ and $V^{\circ}(r)$
are real, symmetric and continuous. In particular cases 
$V^{\circ}_{\alpha\beta}(r) =
V^{\circ}_{\alpha\alpha}(r)\delta_{\alpha\beta}$ or
$V^{\circ}_{\alpha\beta}(r)=0$.
The system (\ref{4.3}) is obtained from Eqs.(\ref{4.1}) multiplied by
$c_{\beta}$, summed over ${\beta}$ at fixed values $\gamma_{\alpha}=
\gamma'_{\alpha}$; $c_{\beta}$ are some spectral characteristics.
For instance, if $\gamma'^{2}=E'$ is the energy of the bound state for
$||V_{\alpha \beta}||$, then elements $c_{\alpha}$ form its normalization
matrix $||C_{\alpha \beta}||=||c_{\alpha}c_{\beta}||$.
We shall search for solutions $\phi_{\alpha\beta}(r)$ of Eq.(\ref{4.1})
with initially unknown potential matrix $V_{\alpha\beta}(r)$ in terms of
solutions appropriate for the known $V^{\circ}_{\alpha\beta}(r)$
in one of the equivalent forms
\begin{eqnarray}
\label{4.4}
\phi_{\alpha\beta}(\gamma,r) &=& \phi^{\circ}_{\alpha\beta}(\gamma,r) -
\psi_{\alpha}(r)\sum_{j}^{N}
W\{\psi^{\circ*}_{j}(r),\phi^{\circ}_{j\beta}(\gamma,r)\}/
(\gamma_{j}^{\prime 2}-\gamma_{j}^{2})\nonumber\\
\phi_{\alpha\beta}(\gamma,r) &=& \phi^{\circ}_{\alpha\beta}(\gamma,r) -
\psi_{\alpha}(r)\sum_{j}^{N}
\int_{0(r)}^{r(b)} h(r')\psi^{\circ*}_{j}(r')
\phi^{\circ}_{j\beta}(\gamma,r')dr'.
\end{eqnarray}
With $\gamma^{2}_{\alpha}=\gamma'^{2}_{\alpha}$, the second expression
(\ref{4.4}) after multiplication by $c_{\beta}$ and summation over
$\beta$ can be rewritten as
\begin{eqnarray}
\label{4.5}
    \psi_{\alpha}(r)={\psi^{\circ}_{\alpha}(r)\over
1+\sum_{j}^{N}\int_{0(r)}^{r(b)} h(r')|\psi^{\circ}(r')|^2}.
\end{eqnarray}
Let us determine the conditions for potential matrix $||V_{\alpha \beta}||$
at which functions $\phi$ and $\psi$ specified by formulas (\ref{4.4}) 
and (\ref{4.5}) will satisfy the system of Eqs.(\ref{4.1}), (\ref{4.3}).
\newpage
Perform double differentiation of relation (\ref{4.4}) and transform
the result with regard to Eqs.(\ref{4.3}) for $\psi_{\alpha}$, (\ref{4.1})
for $\phi^{\circ}_{\alpha\beta}$, and definition (\ref{4.4})
for $\phi_{\alpha\beta}$ and carrying further simplifications we obtain
\begin{eqnarray}
\label{4.6}
-{d^2\phi_{\alpha \beta}(r)\over dr^2} &+&
\sum_{\beta'}^{N} V_{\alpha \beta'}(r)\phi_{\beta'\beta}(r)
- \gamma^{2}_{\alpha}h(r)\phi_{\alpha \beta}(r)=
\sum_{\beta'}^{N} [V_{\alpha \beta'}(r) - V^{\circ}_{\alpha \beta'}(r)]
\phi^{\circ}_{\beta'\beta}\nonumber\\
&+&2{d[h(r)\psi_{\alpha}(r)\sum_{\beta'}^{N}\psi^{\circ*}_{\beta'}(r)]
\over dr}\phi^{\circ}_{\beta'\beta}
+{dh(r)\over dr}\psi_{\alpha}(r)\sum_{\beta'}^{N}\psi^{\circ*}_{\beta'}(r)
\phi^{\circ}_{\beta'\beta}.
\end{eqnarray}
One can easily see that the matrix of functions $\phi_{\alpha\beta}$
satisfy the system (\ref{4.1}) if the right-hand side of Eq.(\ref{4.6})
vanishes. In virtue of linear independence of functions
$\phi^{\circ}_{\alpha\beta}$ this is equivalent to
\begin{eqnarray}
\label{4.7}
V_{\alpha \beta}(r)= V^{\circ}_{\alpha\beta}(r)-
2(d/dr)[h(r)\psi_{\alpha}(r){\psi}^{\circ*}_{\beta}(r)]
+(dh(r)/dr)\psi_{\alpha}(r){\psi}^{\circ*}_{\beta}(r)
\end{eqnarray}
Taking account of definition (\ref{4.5}) for $\psi_{\alpha}(r)$
in Eqs.(\ref{4.7}) and (\ref{4.4}),
we find the analytical relationships between different potential matrices
and their pertinent solutions. Now it is easy to establish analytical
relationships between the solutions for different potential matrices
and the potentials themselves in a more general case being the matrix
generalization of the single-channel problem (\ref{3.1}) -- (\ref{3.8}).

\section{ Conclusions.}
\noindent
Generalization is given for the algebraic transformations for equations
(\ref{1.1}) with a functional dependence $h(r)$ in the right-hand side
of the Schr\"odinger equations. Analytical relationships between
different potential matrices and their pertinent solutions are
constructed that appear to be generalization of the corresponding
single-channel formulae. Under a certain choice of $h(r)$, the Bargmann
and Darboux transformations for both fixed and varying $l$ and $E$ are
particular cases of the generalized transformations.

\nonumsection{Acknowledgements}
\noindent
This work was supported in part by the Foundation of the "Heisenberg-Landau
Program" of investigations.
The author is grateful to Professor H.von Geramb for useful discussions.

\nonumsection{References}
{}
\end{document}